\documentclass[prb, twocolumn, superscriptaddress]{revtex4}
\usepackage{graphicx,color}
\usepackage{amssymb}
\usepackage{amsmath}
\usepackage{mathrsfs}
\usepackage{bm}
\usepackage{hyperref}
\usepackage{tabularx}
\usepackage{multirow}
\usepackage{braket}
\usepackage{xcolor}
\usepackage{gensymb}
\usepackage{makecell}

\DeclareMathOperator*{\bdg}{BdG}

\DeclareMathOperator*{\tr}{Tr}

\begin{document}

\title{Topological Superconductivity in Monolayer T$_{\textrm{d}}$-MoTe$_2$}
\author{Xin-Zhi Li} \thanks{These authors contributed equally to this work}
\affiliation{School of Physical Science and Technology, ShanghaiTech University, Shanghai 201210, China}
\author{Zhen-Bo Qi}	\thanks{These authors contributed equally to this work}
\affiliation{School of Physical Science and Technology, ShanghaiTech University, Shanghai 201210, China}
\author{Quansheng Wu}
\affiliation{Beijing National Laboratory for Condensed Matter Physics and Institute of Physics, Chinese Academy of Sciences, Beijing 100190, China}
\affiliation{University of Chinese Academic of Sciences, Beijing 100049, China}
\author{Wen-Yu He} \thanks{hewy@shanghaitech.edu.cn}
\affiliation{School of Physical Science and Technology, ShanghaiTech University, Shanghai 201210, China}

\date{\today}
\date{\today}

\begin{abstract}
Topological superconductivity has attracted significant attention due to its potential applications in quantum computation, but its experimental realization remains challenging. Recently, monolayer T$_{\textrm{d}}$-MoTe$_2$ was observed to exhibit gate tunable superconductivity, and its in-plane upper critical field exceeds the Pauli limit. Here, we show that an in-plane magnetic field beyond the Pauli limit can drive the superconducting monolayer T$_{\textrm{d}}$-MoTe$_2$ into a topological superconductor. The topological superconductivity arises from the interplay between the in-plane Zeeman coupling and the unique \emph{Ising plus in-plane SOC} in the monolayer T$_{\textrm{d}}$-MoTe$_2$. The \emph{Ising plus in-plane SOC} plays the essential role to enable the effective $p_x+ip_y$ pairing. Importantly, as the essential \emph{Ising plus in-plane SOC} in the monolayer T$_{\textrm{d}}$-MoTe$_2$ is generated by an in-plane polar field, our proposal demonstrates that applying an in-plane magnetic field to a gate tunable 2D superconductor with an in-plane polar axis is a feasible way to realize topological superconductivity. 

\end{abstract}

\maketitle

\section{Introduction}
Topological superconductors hosting unpaired Majorana zero modes are known as a potential platform for topological quantum computation~\cite{Kitaev1, Kitaev2, ReadGreen, Ivanov, Sarma1, Alicea1, Alicea3, Kane1, Shoucheng2, Franz, Sato2}, which have stimulated the intense pursuit of topological superconductivity (TSC) in real materials~\cite{LiangFu1, Xiaoliang, Sato, Sau, Sarma2, Oppen, Alicea2, Patrick1, Shoucheng, EuaKim, Jinfeng, DingHong, Kezilebieke, Yulin, Sau2, Stern, FengLiu}. The key element to realize TSC is to have the $p$-wave pairing in the Fermi surfaces~\cite{Kitaev1, ReadGreen, LiangFu1}, but the intrinsic $p$-wave superconductors are extremely rare and yet to confirm. In the past decades, intensive efforts have been made attempting to induce the effective $p$-wave pairing via the proximity effect with an $s$-wave superconductor~\cite{LiangFu1, Xiaoliang, Sato, Sau, Sarma2, Oppen, Alicea2, Patrick1, Jinfeng, Sau2, Stern}. Among a wide variety of proposals, one ingenious setup is the hybrid superconductor-semiconductor system with strong spin-orbit coupling (SOC) subjected to a Zeeman field~\cite{Sato, Sau, Sarma2, Oppen, Alicea2, Patrick1}, where the Zeeman coupling results in odd number of Fermi surfaces and makes the proximity pairing in one single Fermi surface become the effective $p$-wave pairing. The complexity of the hybrid superconductor-semiconductor structure and its reliance on the pairing proximity effect, however, largely limits the experimental realization of TSC. Recently, several two dimensional (2D) materials are found to exhibit the gate tunable superconductivity ~\cite{Justin1, Justin2, Fai1, Iwasa, Barrera, Fatemi, Sajadi, Caoyuan, Park, Andrea0, Andrea1, Andrea, Park2, ZhangYiran, Holleis, Pantaleon, Faxian, Guangtong, Rhodes, Jindal, Smet, Wakamura, YuanGan}. Therefore, seeking an intrinsic 2D superconductor that is highly tunable to support the TSC phase becomes an alternative approach that bypasses the unfavorable conditions of the hybrid superconductor-semiconductor device.

MoTe$_2$ is a layered transition metal dichaocogenide that exhibits rich topological and superconductivity properties~\cite{Guangtong, Rhodes, Jindal, Smet, Wakamura, YuanGan, Zhijun, Yanpeng, Wudi}. The MoTe$_2$ of T$_{\textrm{d}}$ structure is an intrinsic superconductor in the bulk~\cite{Yanpeng, Wudi}, and the superconductivity gets greatly enhanced as the T$_{\textrm{d}}$-MoTe$_2$ is exfoliated down to the monolayer~\cite{Rhodes, Jindal, Smet, Wakamura, YuanGan}. In the dual-gate device of monolayer T$_{\textrm{d}}$-MoTe$_2$, superconductivity has been achieved in both the electron and hole dominated region~\cite{Rhodes, Smet}. Owing to the gate tunability, the electron and hole superconducting states can be converted to each other~\cite{Rhodes, Smet}. According to the first principle band structure calculations, the Fermi surfaces of monolayer T$_{\textrm{d}}$-MoTe$_2$ are composed of one hole pocket centred at $\Gamma$ along with two flanking electron pockets~\cite{Guangtong, Rhodes, Junwei, Mengli}. Given the condition that the chemical potential is gate tuned to be around the energy of $\Gamma$, an in-plane magnetic field induced Zeeman gap at $\Gamma$ would leave one single hole Fermi surface, which creates the prerequisite to realize the effective $p$-wave pairing in the monolayer T$_{\textrm{d}}$-MoTe$_2$.

The monolayer T$_{\textrm{d}}$-MoTe$_2$ involves a special type of SOC that is a mixture of the Ising SOC and in-plane SOC~\cite{Guangtong, Rhodes}. The SOC arises from the gate induced out of plane electric field~\cite{JustinSong} and the inversion symmetry breaking of the crystal structure~\cite{Suyang}. Importantly, for the hole superconducting states in the monolayer T$_{\textrm{d}}$-MoTe$_2$, the \emph{Ising plus in-plane SOC} resembles the out of plane Dresselhaus SOC and in-plane Rashba SOC in the hybrid superconductor-$\left(110\right)$-grown-semiconductor system~\cite{Alicea2}, which can be driven into a topological superconducting state by applying an in-plane magnetic field. In the superconducting T$_{\textrm{d}}$-MoTe$_2$ of atomic thickness, it has been demonstrated that the \emph{Ising plus in-plane SOC} gives rise to an enhanced in-plane upper critical field with an emergent two fold symmetry~\cite{Guangtong}. Since the Ising SOC component protects the pairing from the in-plane magnetic field~\cite{Justin2, Fai1, Iwasa, Barrera}, it becomes promising that the gate tunable superconducting monolayer T$_{\textrm{d}}$-MoTe$_2$ in an in-plane magnetic field can enter the TSC phase.

Here, we propose that the gate tunable superconducting monolayer T$_{\textrm{d}}$-MoTe$_2$ can be driven into a topological superconductor by applying an in-plane magnetic field. We show that the unique \emph{Ising plus in-plane SOC} in the monolayer T$_{\textrm{d}}$-MoTe$_2$ is essential to the in-plane magnetic field induced TSC phase. For the monolayer T$_{\textrm{d}}$-MoTe$_2$ with a polar axis tilted to its atomic layer~\cite{Guangtong, Jindal, Rhodes, JustinSong}, we point out that for the TSC phase transition, the preferred direction of the in-plane magnetic field is along the in-plane projection of the polar axis. By performing the mean field calculation, we demonstrate that the in-plane magnetic field induced TSC phase occupies a considerable region in the phase diagram of the superconducting monolayer T$_{\textrm{d}}$-MoTe$_2$. As entering the TSC phase is always accompanied by the in-plane magnetic field driven gap closing, the tunneling spectroscopy to measure the density of states (DOS) is suggested to experimentally detect the phase transition to the TSC. It is worth emphasizing the key advantages held in our scheme of TSC realization in the monolayer T$_{\textrm{d}}$-MoTe$_2$: i) the TSC phase is derived from an intrinsic 2D superconductor; ii) the monolayer nature circumvents the undesired orbital depairing effect of the in-plane magnetic field; iii) the monolayer T$_{\textrm{d}}$-MoTe$_2$ is highly tunable through the dual-gate and in-plane magnetic field.

\begin{figure*}[t]
\centering
\includegraphics[width=0.95\textwidth]{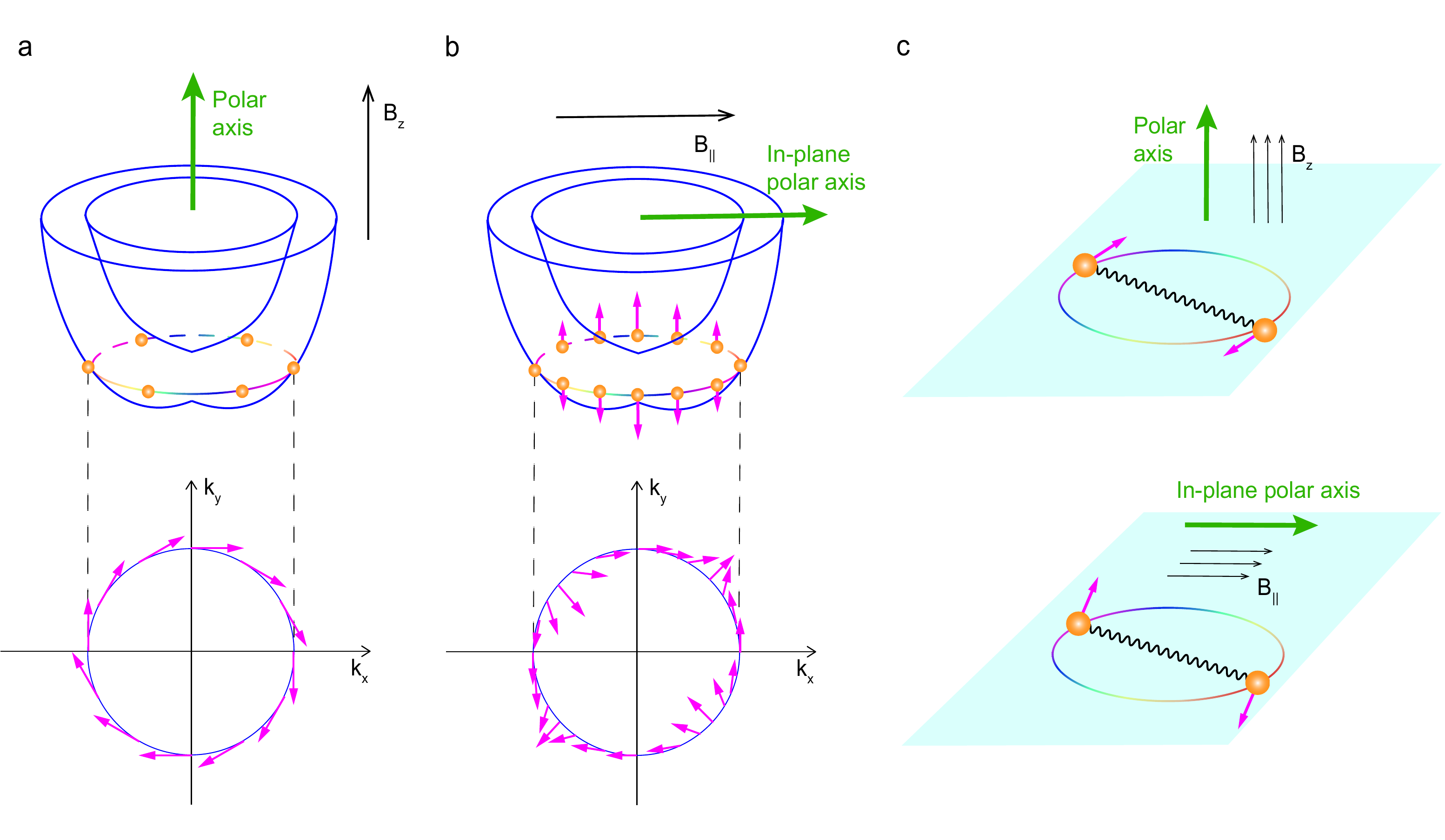}
\caption{The schematic diagram of realizing topological superconductivity in systems with SOC. (a) The schematic band dispersions with Rashba SOC (upper panel) and the Rashba SOC induced spin texture in the Fermi surface (lower panel). The system with Rashba SOC has a $z$-directional polar axis. Applying a magnetic field parallel to the $z$-directional polar axis introduces a Zeeman gap at $\bm{k}=\bm{0}$. (b) The schematic band dispersions  and the corresponding spin textures derived from a 2D system with an in-plane mirror symmetry (the C$_{1v}$ symmetry). The mirror plane contains an in-plane polar axis that lies in the atomic plane. The resulting in-plane polar field generates the unique \emph{Ising plus in-plane} SOC, where its $z$-directional and in-plane components of the spin texture are plotted in the upper and lower panel respectively. Similar to the case of Rashba SOC, an in-plane magnetic field parallel to the in-plane polar axis leads to a Zeeman gap at $\bm{k}=\bm{0}$. (c) In the superconducting state, the effective $p_x+ip_y$ pairing is induced by the magnetic field applied parallel to the polar axis. The upper and lower panel corresponds to the effective $p_x+ip_y$ pairing in the 2D system with the Rashba SOC and the \emph{Ising plus in-plane SOC} respectively.}
\label{fig1}
\end{figure*}

\section{Results}
\subsection{Realization of TSC in systems with SOC}
To illustrate the essential role of the \emph{Ising plus in-plane SOC} in our TSC realization in the monolayer T$_{\textrm{d}}$-MoTe$_2$, we first start with reviewing the seminal scheme of effective $p_x+ip_y$ pairing in the Rashba superconductor~\cite{Sato, Sau, Sarma2, Oppen}. For a crystalline system with SOC, its general Hamiltonian in the Bloch basis $\left[c_{\bm{k},\uparrow}, c_{\bm{k},\downarrow}\right]$ can be written as
\begin{align}\label{SOC_01}
H\left(\bm{k}\right)=\epsilon_{\bm{k}}-\mu+\bm{g}_{\bm{k}}\cdot\bm{\sigma}+\frac{1}{2}g\mu_{\textrm{B}}\bm{B}\cdot\bm{\sigma},
\end{align}
with $\epsilon_{\bm{k}}$ being the band energy in terms of the Bloch wave vector $\bm{k}$, $\mu$ being the chemical potential, $\bm{g}_{\bm{k}}$ denoting the SOC, and $\bm{\sigma}=\left(\sigma_x,\sigma_y,\sigma_z\right)$ being the Pauli matrices in the spin space. The last term in Eq. \ref{SOC_01} represents the magnetic field induced Zeeman coupling,  where $g=2$ is the Land\'e $g$ factor and $\mu_{\textrm{B}}$ is the Bohr magneton. For a 2D electron gas with a $z$-directional polar axis, an electric field along the polar axis generates the Rashba SOC $\bm{g}_{\bm{k}}=\alpha\left(k_y,-k_x, 0\right)$. The Rashba SOC pins the electrons' spins to the in-plane direction and gives a helical spin texture as shown in Fig. \ref{fig1} (a). Given an isotropic band dispersion $\epsilon_{\bm{k}}=\frac{\hbar^2\bm{k}^2}{2m}$, the Zeeman effect of a magnetic field parallel to the $z$-directional polar axis lifts the Kramers degeneracy at $\bm{k}=\bm{0}$ and opens the Zeeman gap. When the chemical potential is tuned into the Zeeman gap, there is only one single Fermi surface present in the system. Crucially, due to the Rashba SOC, the $z$-directional magnetic field cannot fully polarize the electrons' spins to the $z$-direction, so the electrons with canted spins in the single Fermi surface are allowed to have finite singlet pairing~\cite{Sato, Sau, Sarma2, Oppen}. Under the condition of weak pairing $\sqrt{\Delta^2+\mu^2}<\frac{1}{2}g\mu_{\textrm{B}}|\bm{B}|$, the pairing in the single Fermi surface becomes the effective $p_x+ip_y$ pairing, making the Rashba superconductor become a topological superconductor~\cite{Sato, Sau, Sarma2, Oppen}.

\begin{table}[htbp]
    \centering
    \caption{The explicit forms of the SOC $\bm{g}_{\bm{k}}\cdot\bm{\sigma}$ in a quasi-2D system. The quasi-2D system respects the C$_1$, $C_{1v}$, C$_2$ or C$_{2v}$ point group symmetry so that the SOC pseudo-vector $\bm{g}_{\bm{k}}$ has a nonzero $z$ component. The terms $\bm{g}_{\bm{k}}\cdot\bm{\sigma}$ are expanded relative to $\Gamma$ at $\bm{k}=\bm{0}$. The coordinate system used in the expansion is fixed by the in-plane polar axis direction.}
    \begin{tabular}{ccc}
        \hline
        \hline
        Point group & $\bm{g}_{\bm{k}}\cdot\bm{\sigma}$ & In-plane polar axis\\ [0.5ex]
        \hline
        C$_1$ & $\sum_{i=x,y,z;j=x,y}\lambda_{i,j}\sigma_ik_j$ & arbitrary\\ [0.5ex]
        C$_{1v}$ & $\lambda_x\sigma_xk_y+\lambda_y\sigma_yk_x+\lambda_z\sigma_zk_y$ & $x$\\ [0.5ex]
        C$_2$ & $\lambda_x\sigma_xk_x+\lambda_y\sigma_yk_y+\lambda_z\sigma_zk_y$ & $x$\\ [0.5ex]
		C$_{2v}$ & $\lambda_z\sigma_zk_y$ & $x$\\[0.5ex]
        \hline
    \end{tabular}\label{tb1}
\end{table}

One long standing challenge for the TSC realization in the Rashba superconductor is known to be the orbital depairing effects introduced by the $z-$directional magnetic field~\cite{Alicea2}. A modification to the Rashba superconductor scheme is to introduce an extra out of plane Dresselhaus SOC so that the transition to the TSC phase can be induced by an in-plane magnetic field~\cite{Alicea2}. However, an intrinsic 2D superconductor with such in-plane and out of plane SOC that enables the in-plane magnetic field driven topological phase transition remains yet to know. Microscopically, the SOC arises from the relativistic interaction between the electron's spin and the crystal electric field $\bm{E}\left(\bm{r}\right)$. The approximate form of the SOC field takes $\bm{g}_{\bm{k}}\propto\bm{E}\left(\bm{r}\right)\times\left(\hbar\bm{k}+\hat{\bm{p}}\right)$ with $\hat{\bm{p}}$ being the momentum operator~\cite{Winkler}. For the 2D Rashba superconductor, it is clear that its Rashba SOC is produced by the $z$-directional polar field. In a 2D system, the $z$-directional SOC that resembles the out of plane Dresselhaus SOC can only be generated by an in-plane polar field. As a result, in a 2D system, a polar axis with in-plane projection is required for the $z$-directional SOC. Such in-plane projection of the polar axis is referred as the in-plane polar axis.

In the crystalline system, the SOC field $\bm{g}_{\bm{k}}$ is known to respect the symmetry constraint~\cite{Samokhin, Agterberg}
\begin{align}
\bm{g}_{\bm{k}}=\det\left(\hat{R}\right)\hat{R}\bm{g}_{\hat{R}^{-1}\bm{k}},
\end{align}
with $\hat{R}$ being the orthogonal matrix that represents the point group transformation. In the quasi-2D case of atomic thickness, it has been pointed out that only C$_1$, C$_{1v}$, C$_{2}$ and C$_{2v}$ symmetries can have the in-plane polar axis and allow the nonzero $z$-component of $\bm{g}_{\bm{k}}$ up to the linear order of $\bm{k}$~\cite{wenyu1}. The explicit forms of $\bm{g}_{\bm{k}}$ expanded from $\Gamma$ are listed in Table \ref{tb1}. It is clear from Table \ref{tb1} that $\bm{g}_{\bm{k}}$ derived from the C$_1$, C$_{1v}$ or C$_{2}$ symmetry is the \emph{Ising plus in-plane SOC} that involves both the $z$-directional and in-plane components, while that from the C$_{2v}$ symmetry is the pure Ising SOC~\cite{Tongzhou} with only the $z$ component of $\bm{g}_{\bm{k}}$ nonzero. Superconductors with the pure Ising SOC can have the nodal TSC phase driven by an in-plane magnetic field~\cite{wenyu2}. Since the mechanism of the effective $p_x+ip_y$ pairing in the Rashba superconductor tells that the realization of the fully gapped TSC phase requires $\bm{g}_{\bm{k}}$ to have at least two nonzero components, the $\bm{g}_{\bm{k}}$ in the quasi-2D system with the C$_1$, C$_{1v}$ or C$_{2}$ symmetry meets the criteria. As a result, the seek for the in-plane magnetic field driven TSC phase focuses on the intrinsic 2D superconductivity with the C$_1$, C$_{1v}$ or C$_{2}$ symmetry. 

Importantly, the \emph{Ising plus in-plane SOC} in the quasi-2D system with the C$_{1v}$ symmetry, as can be seen from the spin texture of $\bm{g}_{\bm{k}}=\left(\lambda_xk_y,\lambda_yk_x,\lambda_zk_y\right)$ in Fig. \ref{fig1} (b), is very similar to the combination of the out of plane Dresselhaus SOC and in-plane Rashba SOC in the hybrid superconductor-(110)-grown-semiconductor system~\cite{Alicea2}. The C$_{1v}$ group only has one in-plane mirror symmetry, so the in-plane polar axis is defined along the intersection of the mirror plane and the layer plane. For an intrinsic 2D superconductor with C$_{1v}$ symmetry that can have its chemical potential gate tuned to be around the energy of $\Gamma$, it is expected that applying an in-plane magnetic field parallel to the in-plane polar axis can lead to the TSC phase in the similar mechanism as the hybrid superconductor-(110)-grown-semiconductor system~\cite{Alicea2}. As is schematically shown in Fig. \ref{fig1} (c), the in-plane magnetic field parallel to the in-plane polar axis induces the effective $p_x+ip_y$ pairing, similar to the effect of the $z$-directional magnetic field parallel to the $z$-directional polar axis in the Rashba superconductor. Since it is promising to engineer a 2D superconductor with the C$_{1v}$ symmetry into a topological superconductor, the superconducting monolayer T$_{\textrm{d}}$-MoTe$_2$ that belongs to the C$_{1v}$ point group turns out to be a suitable candidate for TSC.


\begin{figure}[t]
\centering
\includegraphics[width=0.5\textwidth]{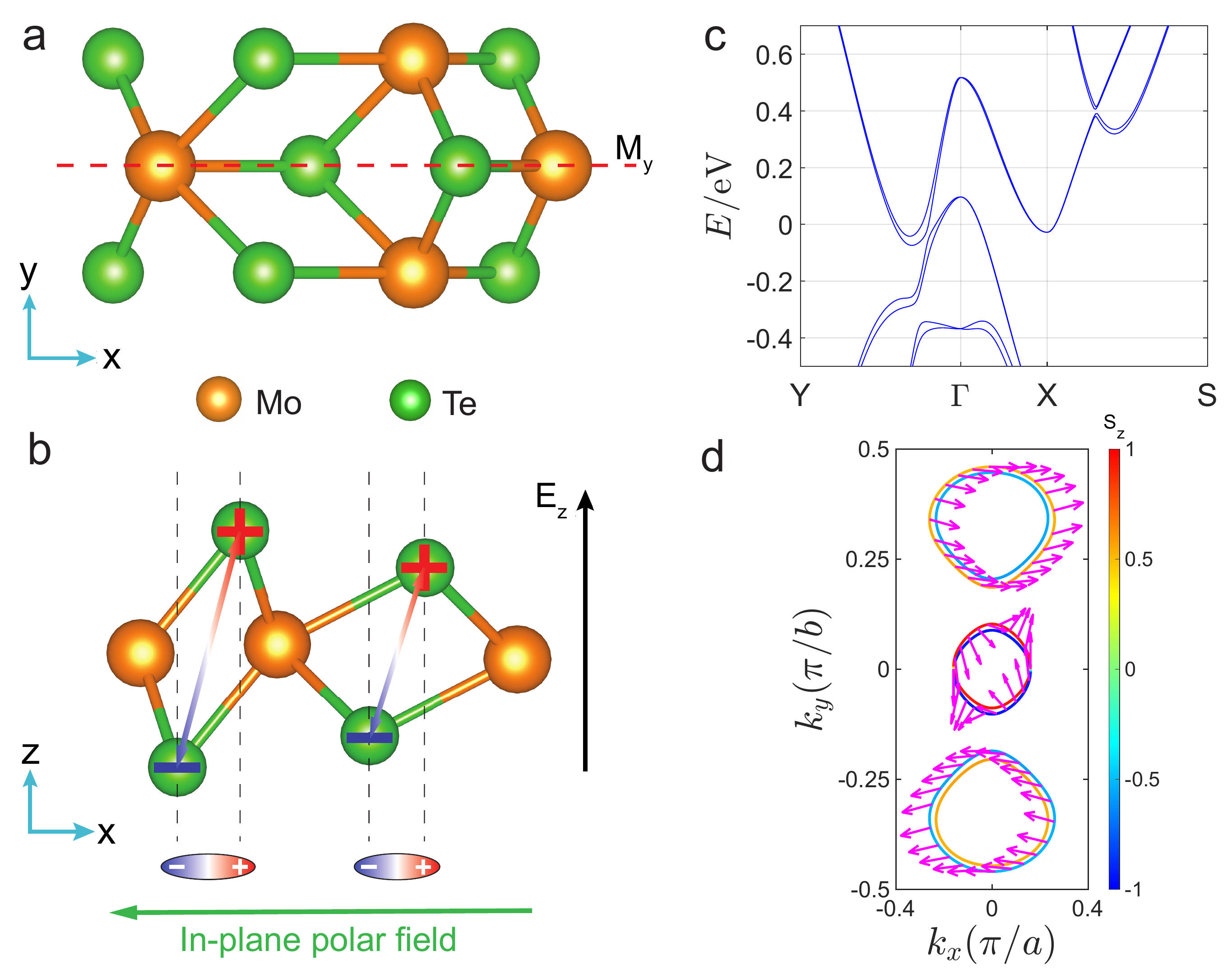}
\caption{The crystal structure and electronic band structure of monolayer T$_d$-MoTe$_2$. (a) The top view of the crystal structure for monolayer T$_d$-MoTe$_2$. The red dashed line labeled by $M_y$ denotes the in-plane mirror symmetry. (b) Due to the nonalignment of the Te atoms in the vertical direction, an out-of-plane electric field $E_z$ induces a net in-plane polar field. (c) The electronic band dispersions along Y-$\Gamma$-X-S for the monolayer T$_d$-MoTe$_2$. A $z$-directional electric field $E_z=0.1~\textrm{V/\AA}$ is considered in the DFT calculations. (d) The spin textures of the $\Gamma$ centered hole pocket and the two flanking electronic pockets.}
\label{fig2}
\end{figure}

\subsection{Electronic structure of the monolayer T$_{\textrm{d}}$-MoTe$_2$}
The crystal structure of the monolayer T$_{\textrm{d}}$-MoTe$_2$ hosts only one in-plane mirror symmetry along $x$ as is shown in Fig. \ref{fig2} (a). Isolated from the inversion symmetry breaking bulk T$_{\textrm{d}}$-MoTe$_2$, the inversion symmetry of the monolayer T$_{\textrm{d}}$-MoTe$_2$ gets slightly broken~\cite{Rhodes, Jindal, Smet, Suyang}. In the dual-gate device, an out of plane electric field further enhances the inversion symmetry breaking~\cite{JustinSong}: the applied $E_z$ field induces a tilted dipole moment as illustrated in Fig. \ref{fig2} (b). The in-plane component of the $E_z$ field induced dipole moment gives rise to the $z$-directional Ising SOC, while its out of plane component generates the in-plane SOC. Under the assumption of the applied $E_z=0.1~\textrm{V/\AA}$, we perform the first principle density functional theory (DFT) calculation to obtain the band structure of the monolayer T$_{\textrm{d}}$-MoTe$_2$ (see the methods section). The results are depicted in Fig. \ref{fig2} (c). Consistent with the previous calculations~\cite{Guangtong, Rhodes, Junwei, Mengli}, the Fermi surfaces are composed of one hole pocket at $\Gamma$ and two electron pockets at either side of $\Gamma$. Such metallic band structure of monolayer T$_{\textrm{d}}$-MoTe$_2$ is supported by the recent transport measurements~\cite{Rhodes, Pawlik}. Deviating from $\Gamma$, spin splittings due to the SOC are observed to appear. On the Fermi surfaces, the \emph{Ising plus in-plane SOC} is clearly demonstrated by the spin textures plotted in Fig. \ref{fig2} (d). 


Since the TSC phase arises from the effective $p_x+ip_y$ pairing in the Fermi surface that encloses the time reversal invariant momenta~\cite{Sato, Sau, Sarma2, Oppen, Alicea2}, our primary focus is on the hole pocket centred at $\Gamma$. In the Bloch basis of $\left[c_{\bm{k},\uparrow}, c_{\bm{k},\downarrow}\right]$ near $\Gamma$, an effective Hamiltonian respecting the C$_{1v}$ symmetry can be constructed to describe the hole bands as
\begin{align}\nonumber\label{H0_MoTe2}
H_0\left(\bm{k}\right)=&-\frac{\hbar^2k_x^2}{2m_x}-\frac{\hbar^2k_y^2}{2m_y}-\mu\\
&+\lambda_xk_y\sigma_x+\lambda_yk_x\sigma_y+\lambda_zk_y\sigma_z,
\end{align}
with $m_x=0.75m_e$, $m_y= 1.07m_e$ being the effective mass (here $m_e$ is the electron mass), and $\lambda_x=2.94\times 10^{-2}${eV$\cdot$\AA}, $\lambda_y=4.13\times 10^{-2}${eV$\cdot$\AA}, $\lambda_z=7.50\times 10^{-2}${eV$\cdot$\AA} denoting the strength of SOC. In experiment, the SOC is tunable by the gate induced $E_z$ field. In the experimental accessible range of $E_z\in\left[-0.1, 0.1\right]\textrm{V/\AA}$~\cite{Jindal,ZhangYiran, Holleis}, our first principle DFT calculations show that the \emph{Ising plus in-plane SOC} in the effective Hamiltonian in Eq. \ref{H0_MoTe2} is sufficient to capture the spin texture in the hole pocket (see the Supplementary Note 1 and Supplementary Note 2). 

\subsection{TSC realization in the monolayer T$_{\textrm{d}}$-MoTe$_2$}
The monolayer T$_{\textrm{d}}$-MoTe$_2$ is found to enter the superconductivity phase with the transition temperature being 5K$\sim$8K~\cite{Rhodes, Jindal, Smet}. Distinctly different from its isostructural monolayer T$_{\textrm{d}}$-WTe$_2$ where the superconductivity occurs only in the electron type carriers~\cite{Fatemi, Sajadi}, the monolayer T$_{\textrm{d}}$-MoTe$_2$ hosts intrinsic superconductivity in both the electron and hole dominated region~\cite{Rhodes, Jindal, Smet}, indicating that there exists finite pairing in the hole pocket at $\Gamma$. Importantly, due to the protection of the Ising SOC component, the in-plane upper critical field of the superconducting monolayer T$_{\textrm{d}}$-MoTe$_2$ gets strongly enhanced: for the hole superconducting states with $T_{\textrm{c}}\approx 7$K, the measured in-plane upper critical field $B_{\textrm{c}2}$ is extrapolated to reach $B_{\textrm{c}2}/B_{\textrm{P}}\approx 2\sim3$~\cite{Rhodes, Smet}, where $B_{\textrm{P}}\approx1.84 T_{\textrm{c}}$ is the Pauli paramagnetic limit~\cite{Clogston, Chandrasekhar}. Since the monolayer T$_{\textrm{d}}$-MoTe$_2$ has an atomic thickness of 0.35 nm that is far smaller than the measured coherence length of $8\sim16$ nm~\cite{Rhodes, Smet}, the orbital magnetic field effect gets suppressed and the Zeeman coupling with the superconducting states plays the dominate role. In the Nambu basis of $\left[c_{\bm{k},\uparrow},c_{\bm{k},\downarrow},c^\dagger_{-\bm{k},\uparrow},c^\dagger_{-\bm{k},\downarrow}\right]$, the Bogliubov-de Gennes (BdG) Hamiltonian for the hole pocket pairing in the presence of an $x$-directional Zeeman coupling takes the form:
\begin{align}\nonumber\label{H_BdG}
H_{\textrm{BdG}}\left(\bm{k}\right)=&\left(-\frac{\hbar^2k_x^2}{2m_x}-\frac{\hbar^2k_y^2}{2m_y}-\mu\right)\tau_z+\frac{1}{2}g\mu_{\textrm{B}}B\tau_z\sigma_x\\
&+\lambda_xk_y\sigma_x+\lambda_yk_x\tau_z\sigma_y+\lambda_zk_y\sigma_z+\Delta\tau_y\sigma_y,
\end{align}
where $\tau_z$ is the Pauli matrix in the particle-hole space and $\Delta$ denotes the pairing order parameter. Similar to the Rashba superconductor, the superconducting state in Eq. \ref{H_BdG} undergoes the topological phase transition and enters the TSC phase for $\frac{1}{2}g\mu_{\textrm{B}}|B|>\sqrt{\Delta^2+\mu^2}$. Specifically, at $\mu=0$, the topological phase transition is estimated to occur around $B=\sqrt{2}B_{\textrm{P}}$~\cite{wenyu2}, which is smaller than the in-plane $B_{\textrm{c}2}$ measured in the superconducting monolayer T$_{\textrm{d}}$-MoTe$_2$~\cite{Rhodes, Smet}. It indicates that the hole pocket pairing in the superconducting monolayer T$_{\textrm{d}}$-MoTe$_2$ can survive to become the effective $p_x+ip_y$ pairing after the in-plane magnetic field driven topological phase transition.

\begin{figure}[t]
\includegraphics[width=0.5\textwidth]{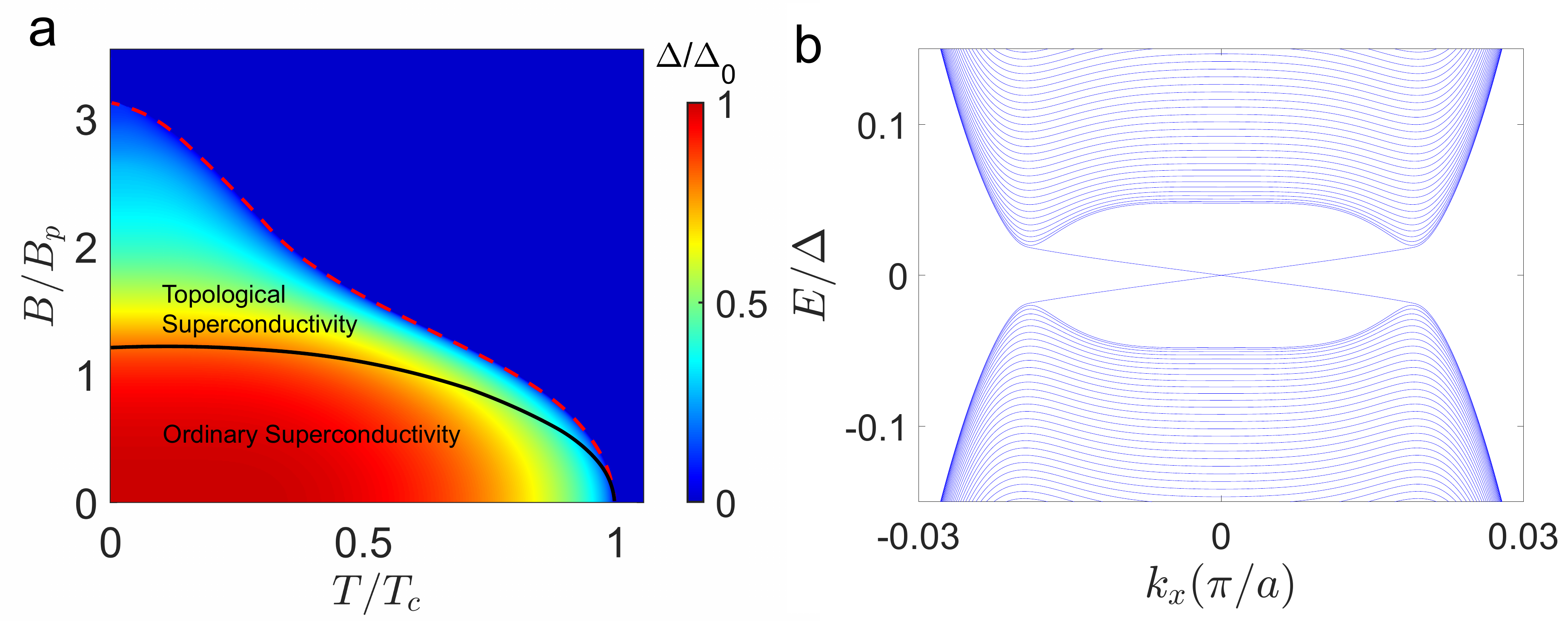}
\caption{The phase diagram of monolayer T$_d$-MoTe$_2$ in an $x$-directional magnetic field. In the region between the in-plane upper critical field $B_{\textrm{c}2}\left(T\right)$ (red dashed line) and the topological phase transition line (black solid line) in (a), the superconducting monolayer T$_d$-MoTe$_2$ becomes a topological superconductor. (b) The BdG quasiparticle energy spectrum of a monolayer T$_d$-MoTe$_2$ ribbon along $x$. The applied $x$-directional magnetic field is $B=2.1\Delta/\left(g\mu_{\textrm{B}}\right)$.}
\label{fig3}
\end{figure}

The above analysis of the in-plane magnetic field driven TSC phase based on the effective Hamiltonian of the $\Gamma$ pocket gets further supported by the mean field phase diagram calculation. At a given temperature $T$, the evolution of the superconducting pairing order parameter $\Delta$ with an applied in-plane magnetic field can be determined by minimizing the free energy density (see the methods section)
\begin{align}\label{BdG_free}
F=&\frac{|\Delta|^2}{U}-\frac{1}{2\beta\Omega}\sum_{\bm{k},\nu}\log\left(1+e^{-\beta\xi_{\nu,\bm{k}}}\right),
\end{align}
with $U$, $\beta=1/\left(k_{\textrm{B}}T\right)$ and $\Omega$ denoting the intra-band attractive interaction, the thermodynamic beta and the volume of the system respectively. Here $\xi_{\nu,\bm{k}}$ are the eigenvalues of the BdG Hamiltonian that is derived from a twelve-band tight binding model of the monolayer T$_{\textrm{d}}$-MoTe$_2$ (see the methods section). In the calculation, the intra-band attractive interaction $U$ is determined by fitting the superconducting critical temperature at zero magnetic field to be 7.6K. The chemical potential $\mu$ is fixed to be at the energy of $\Gamma$ that locates at the top of the hole bands. The resulting $B$-$T$ phase diagram of the superconducting state is shown in Fig. \ref{fig3} (a). It clearly shows that the in-plane $B_{\textrm{c}2}$ is enhanced to reach 3$B_{\textrm{P}}$. By solving $|\Delta\left(T, B\right)|=\frac{1}{2}g\mu_{\textrm{B}}|B|$, we obtain the topological phase transition line. Between the in-plane $B_{\textrm{c}2}$ and the topological phase transition line, it can be seen that there exists a considerable superconducting region supporting the TSC phase. At $T=0$K, given an in-plane magnetic field in the TSC region, the BdG quasiparticle energy spectrum is plotted in Fig. \ref{fig3} (b), where a chiral Majorana edge state appears as expected for the TSC phase.

In reality, the gate tunability in the superconducting monolayer T$_{\textrm{d}}$-MoTe$_2$ facilitates our proposal of the in-plane magnetic field driven TSC. First, the gate tunability enables the conversion of the superconducting state between the electron type and the hole type~\cite{Rhodes, Smet}, indicating that the gate can tune the chemical potential close to the energy of $\Gamma$ at the top of the hole bands. In the presence of an in-plane magnetic field, opening a Zeeman gap at $\Gamma$ reduces the DOS, so the magneto-conductance exhibits a dip once the chemical potential is tuned to lie within the Zeeman gap (see the Supplementary Note 2 and Supplementary Fig. 2). Second, the gate induced $E_z$ field can tune the strength of the SOC. For the \emph{Ising plus in-plane SOC} $\bm{g}_{\bm{k}}=\left(\lambda_xk_y,\lambda_yk_x,\lambda_zk_y\right)$, its $y$ and $z$ components that are perpendicular to the $x$-directional magnetic field play the major role to enhance the in-plane $B_{\textrm{c}2}$ and promote the effective $p_x+ip_y$ pairing, while its $x$ component suppresses the pairing in the presence of an $x$-directional magnetic field. By varying $E_z$ in our first principle DFT calculation, we show that $E_z\in\pm[0.05,0.07]\textrm{V/\AA}$ can reduce the $x$ component of $\bm{g}_{\bm{k}}$ so that the in-plane $B_{\textrm{c}2}$ can be further enhanced and the TSC region gets enlarged (see the Supplementary Note 2 and Supplementary Fig. 3). For the superconducting monolayer T$_{\textrm{d}}$-MoTe$_2$ with dual-gate, the chemical potential, the \emph{Ising plus in-plane SOC}, and the in-plane magnetic field are all individually tunable. Such high tunability makes the superconducting monolayer T$_{\textrm{d}}$-MoTe$_2$ a promising platform to realize the long sought topological superconductivity. 

Given a 2D TSC phase in the monolayer T$_{\textrm{d}}$-MoTe$_2$, one can further cut it into a quasi-1D wire with its width smaller than the decay length of the Majorana edge state. The decay length of the Majorana edge state is estimated to be about 300 nm for the pristine topological superconducting monolayer T$_{\textrm{d}}$-MoTe$_2$ at $B=2.1\Delta/\left(g\mu_{\textrm{B}}\right)$. In such a quasi-1D superconducting wire, a pair of robust Majorana zero modes live in the ends of the wire~\cite{Potter01}, indicating that the quasi-1D wire of topological superconducting monolayer T$_{\textrm{d}}$-MoTe$_2$ can serve as the building block for the Majorana based topological quantum computation~\cite{Sarma1, Alicea1, Alicea3}.

\subsection{Detection}
\begin{figure}[t]
\centering
\includegraphics[width=0.5\textwidth]{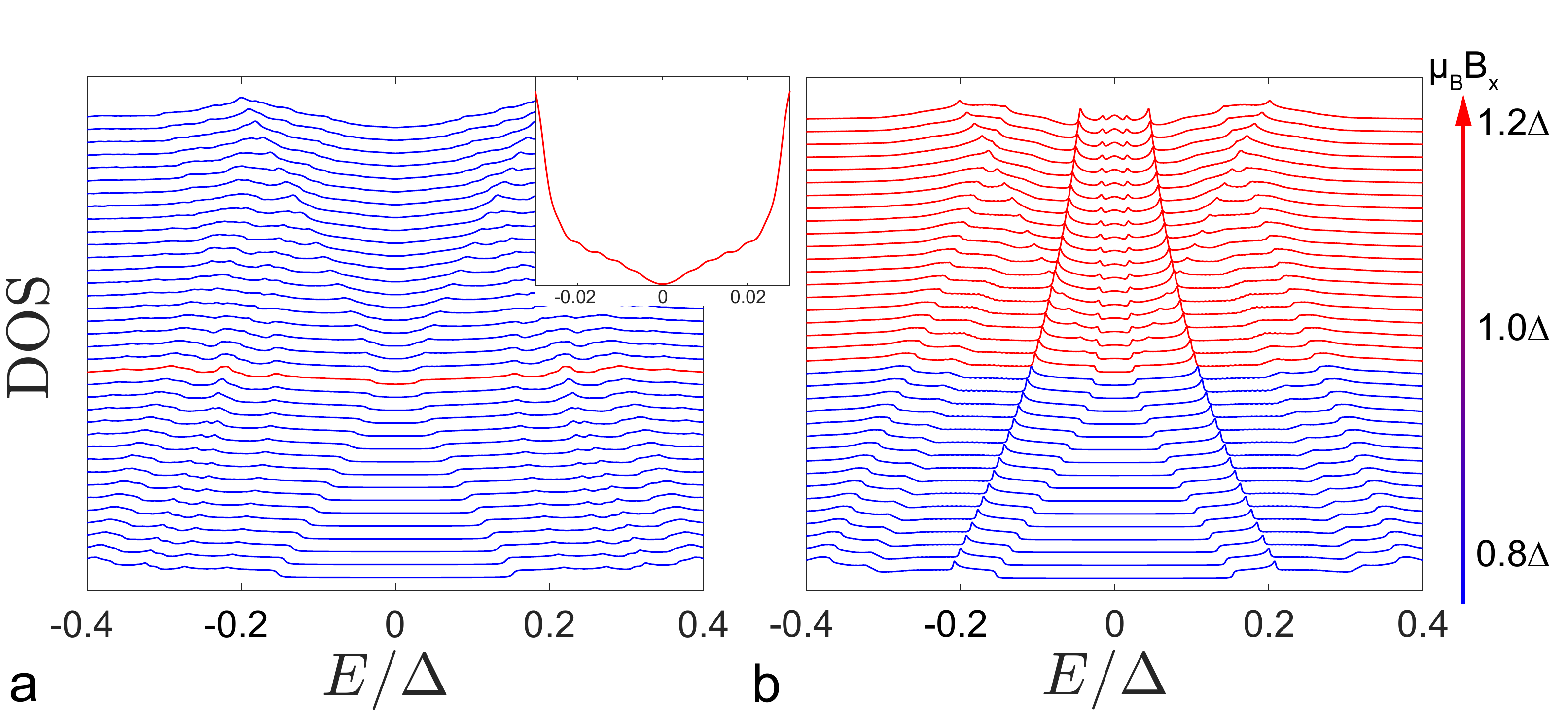}
\caption{The signatures of the in-plane magnetic field driven superconducting topological phase transition. (a) The electronic DOS in the bulk at different in-plane magnetic fields. In the magnetic field driven topological phase transition, the electronic DOS shows the gap closing and reopening. At the topological phase transition with the gap closing at $\bm{k}=\bm{0}$, the DOS exhibits the V-shape (red colored) as shown in the inset. During the topological phase transition, the electronic DOS change from the U-shape to the V-shape and then back to the U-shape. (b) The surface electronic DOS calculated at the edge along $x$ direction. After the topological phase transition, a plateau centered at the zero energy emerges, which corresponds to the chiral Majorana edge states with the linear energy dispersion in the topological superconductivity. The surface electronic DOS after the topological phase transition are colored in red.}
\label{fig4}
\end{figure}

For our proposed in-plane magnetic field driven TSC phase in the superconducting monolayer T$_{\textrm{d}}$-MoTe$_2$, we show that one can measure the evolution of DOS with $\bm{B}$ to identify the topological phase transition. It is known that the topological phase transition must undergo the gap close and reopen in the BdG quasiparticle energy spectrum. Given a small $\bm{B}$, the gap in the BdG energy spectrum is not closed yet, so the electronic DOS exhibits a U-shape with zero DOS in the gap. At the critical value of $\bm{B}$ that enables the topological phase transition, the BdG quasiparticle has the Dirac cone like dispersion with the gap closing at $\bm{k}=\bm{0}$, so the resulting DOS becomes the V-shape. After the topological phase transition, the gap reopens so the electronic DOS turns back to the U-shape. In Fig. \ref{fig4} (a), we present the local electronic DOS in the bulk sample of superconducting monolayer T$_{\textrm{d}}$-MoTe$_2$. Here the electronic DOS is calculated by $N\left(E\right)=-\frac{1}{\pi}\textrm{Im}G^{\textrm{R}}\left(E\right)$ with $G^{\textrm{R}}\left(E\right)$ being the retarded Green's function (see the Supplementary Note 3). Clearly, the electronic DOS evolution in Fig. \ref{fig4} (a) shows the gap close and reopen, consistent with our expected signature of the in-plane magnetic field induced topological phase transition.

After the topological phase transition, the key feature of the TSC phase is known to be the chiral Majorana edge state that appears within the gap. At the edges of the sample, the local electronic DOS in the gap is expected to get strongly enhanced by the chiral Majorana edge states. By calculating the local electronic DOS at the edge of the supeconducting monolayer T$_{\textrm{d}}$-MoTe$_2$ (see the Supplementary Note 3), we find that after the topological phase transition, a plateau in the DOS emerges within the gap of the BdG quasiparticle energy spectrum as shown in Fig. \ref{fig4} (b). The plateau appearing within the gap signifies the chiral Majorana edge states with the linear energy dispersion in the TSC phase. Since both the local electronic DOS in the bulk and at the edge can be detected by the tunneling spectroscopy~\cite{Bruus, wenyu3, Costanzo}, we suggest to carry out tunneling experiments to probe the in-plane magnetic field driven TSC phase.

\section{Discussion}
In the above sections, we have proposed that applying an in-plane magnetic field to the superconducting monolayer T$_{\textrm{d}}$-MoTe$_2$ is promising to achieve the TSC phase. In the proposal, the \emph{Ising plus in-plane SOC} plays the crucial role in the TSC realization. The Ising SOC component is responsible for the enhancement of the in-plane $B_{\textrm{c}2}$ to guarantee the finite superconducting pairing after the magnetic field driven topological phase transition. It is evident that a larger Ising SOC component is more favorable for the TSC realization. For the moderate Ising SOC component in the monolayer T$_{\textrm{d}}$-MoTe$_2$, one can consider to put it in contact with another 2D material with strong Ising SOC so that the Ising SOC component in the monolayer T$_{\textrm{d}}$-MoTe$_2$ can get proximity-enhanced~\cite{ZhangYiran, Holleis, Wakatsuki}. Since the Ising SOC arises from the in-plane polar field, it is also possible to add uniaxial strain to increase the Ising SOC strength~\cite{Manzeli, Martin}. With the Ising SOC component enhanced, the TSC region is expected to get further enlarged in the phase diagram of the monolayer T$_{\textrm{d}}$-MoTe$_2$.

In our proposal, for the \emph{Ising plus in-plane SOC} $\bm{g}_{\bm{k}}=\left(\lambda_xk_y,\lambda_yk_x,\lambda_zk_y\right)$ in the monolayer T$_{\textrm{d}}$-MoTe$_2$, we suggest to apply the in-plane magnetic field along the in-plane mirror axis, namely the $x$ direction, to realize the TSC phase. In the case of applying an in-plane magnetic field along the $y$ direction, the whole system respects the in-plane mirror symmetry, so the $y$-directional magnetic field can only lead to the mirror symmetry protected nodal point degeneracies at $k_y=0$ plane. It indicates that the fully gapped TSC phase requires the applied in-plane magnetic field to have a nonzero $x$-directional component. For the applied in-plane magnetic field slightly deviating from the $x$ direction, the magnetic field can still induce the topological phase transition, but the resulting topological BdG energy spectrum is a bit tilted in the $k_x$ direction (see the Supplementary Note 3 and Supplementary Fig. 5). The tilted BdG energy spectrum indicates that the magnetic field makes the centre of the Fermi surface get shifted away from $\bm{k}=\bm{0}$, so the pairing in the Fermi surface is possible to gain finite momentum under appropriate conditions~\cite{WeiYi, Chuanwei, Yuanfanqi}.

The crucial element in our proposal is the \emph{Ising plus in-plane SOC} in a gate tunable 2D superconductor. For the monolayer T$_{\textrm{d}}$-MoTe$_2$, its \emph{Ising plus in-plane SOC} arises from the C$_{1v}$ symmetry. Apart from the monolayer T$_{\textrm{d}}$-MoTe$_2$, the bilayer T$_{\textrm{d}}$-MoTe$_2$ also belongs to the C$_{1v}$ point group and exhibits gate tunable superconductivity~\cite{Guangtong}. The Fermi surfaces of the bilayer T$_{\textrm{d}}$-MoTe$_2$ include two hole pockets at $\Gamma$~\cite{Guangtong, Jindal}, so it is possible that applying an in-plane magnetic field to the superconducting bilayer T$_{\textrm{d}}$-MoTe$_2$ gives rise to a topological superconductor with the Chern number $C=2$. The seek for the intrinsic 2D superconductors with \emph{Ising plus in-plane SOC} can extend beyond the 4 point groups listed in Table \ref{tb1}. For example, the monolayer 2M-WS$_2$ is a superconductor whose crystal structure respects the C$_{2h}$ symmetry~\cite{Yulin}. In the presence of gate induced out of plane electric field, the C$_{2h}$ symmetry is reduced to the C$_{1v}$ symmetry so that a gate tunable \emph{Ising plus in-plane SOC} can be generated in monolayer 2M-WS$_2$. Similar to our proposal of TSC realization in monolayer T$_{\textrm{d}}$-MoTe$_2$, we expect that an in-plane magnetic field can induce the TSC in monolayer 2M-WS$_2$ as well.

In a brief summary, we propose that for the gate tunable superconducting monolayer T$_{\textrm{d}}$-MoTe$_2$ with the C$_{1v}$ symmetry, the TSC phase can be realized by applying an in-plane magnetic field. Based on the symmetry analysis, we point out that 2D superconductors with an in-plane polar axis are good platforms to host the TSC driven by an in-plane magnetic field.

\section{Methods}

\subsection{The First principle calculation and effective tight binding model}

The band structure of monolayer T$_{\textrm{d}}$-MoTe$_2$ is calculated by first-principle DFT as implemented in the Vienna Ab initio Simulation Package (VASP) using the projected-augmented wave method (PAW)~\cite{VASP_PAW}. We adopt the Perdew-Burke-Ernzerhof's (PBE) form of exchange-correlation functional within the generalized-gradient approximation (GGA). The Brillouin zone is sampled in a Monkhorst-Pack k-point of $9\times 15\times 1$ and a large vacuum slab of 30 $\textrm{\AA}$ along the $z$ direction is used to reduce image interactions under periodic boundary condition in the calculation. The monolayer crystal structure is fully relaxed with a maximum residual force of less than 0.01 $\textrm{eV/\AA}$. The energy cutoff of the plane wave basis is set to be 520 eV. To take into account the dual-gate induced $z$-directional electric field, we have considered an out of plane electric field in the range of {[-0.1, 0.1]V/\AA} in the calculation.

After the first principle DFT calculation is done, a twelve-band tight binding model of monolayer T$_{\textrm{d}}$-MoTe$_2$ is built through Wannier function construction by using the Wannier 90 package~\cite{wannier90}. At the Fermi level around the $\Gamma$ pocket, the dominant orbitals are the $d_{x^2-y^2}$ type from the Mo atom and $p_y$ type from the Te atom. We project twelve bands near the Fermi level into these atomic orbitals. By a Wannier interpolation~\cite{wannier90} of these twelve DFT bands, we construct a basis of localized Wannier functions, comprising four $d_{x^2-y^2}$-type orbitals that are derived from the Mo atoms and eight $p_y$-type orbitals derived from Te atoms. The spin degrees are already involved in the twelve orbitals. The hopping matrix of the twelve-band tight binding Hamiltonian is obtained in this way. After the Fourier transformation, we get the twelve-band tight binding Hamiltonian of the monolayer T$_{\textrm{d}}$-MoTe$_2$ in the $\bm{k}$ space.

\subsection{The mean field free energy calculation for the superconducting pairing state}
Given the twelve-band tight binding model of monolayer T$_{\textrm{d}}$-MoTe$_2$ in the Wannier basis, the Hamiltonian in the presence of an effective pairing attractive interaction can be written as
\begin{align}\nonumber
\mathcal{H}=&\sum_{\bm{k},i,o,o', s, s'}c^{\dagger}_{\bm{k},o,s}[h_{os,o's'}(\bm{k})-\mu+\frac{1}{2}g\mu_{\textrm{B}}B_i\sigma_{i,ss'}]c_{\bm{k},o',s'}\nonumber\\
&-\dfrac{U}{\Omega}\sum_{\bm{k},\bm{k}'}\sum_{o,o'}c^{\dagger}_{\bm{k},o,\uparrow}c^{\dagger}_{-\bm{k},o,\downarrow}c_{-\bm{k}',o',\downarrow}c_{\bm{k}',o',\uparrow},
\end{align}
where $c^{\dagger}_{\bm{k},o,\sigma}$ is the creation operator and $h_{os,o's'}\left(\bm{k}\right)$ denotes the hopping matrix. Here $o$ is the orbital index, $s=\uparrow/\downarrow$ represents the spin index, and $i=x, y, z$ denotes the spatial components of magnetic field. In the mean field approximation, the partition function for the superconducting pairing state can be obtained through the path integral formalism~\cite{Altmanbook}
\begin{align}\nonumber
\mathcal{Z}\approx&\int \mathcal{D}\left[\psi^{\dagger}_{\bm{k}},\psi_{\bm{k}}\right]e^{-\frac{\beta\Omega|\Delta|^2}{U}+\frac{1}{2}\sum_{\bm{k},n}\psi^{\dagger}_{\bm{k},n}\left[i\omega_n-\mathcal{H}_{\bdg}(\bm{k})\right]\psi_{\bm{k},n}}\nonumber\\
=&e^{-\frac{\beta|\Omega|^2}{U}+\frac{1}{2}\sum_{\bm{k},n}\tr\log\left[-\beta G^{-1}(\bm{k},i\omega_n)\right]},
\end{align}
where $\psi^{\dagger}_{\bm{k}}=\left[c^{\dagger}_{\bm{k},o,\uparrow},c^{\dagger}_{\bm{k},o,\downarrow},c_{-\bm{k},o,\uparrow},c_{-\bm{k},o,\downarrow}\right]^{\textrm{T}}$ is the Nambu spinor, $G^{-1}\left(i\omega_n,\bm{k}\right)=i\omega_n-h_{\textrm{BdG}}\left(\bm{k}\right)$ is the Matsubara Green's function with the BdG Hamitonian taking the form
\begin{align}\label{h_BdG}
h_{\textrm{BdG}}\left(\bm{k}\right)=\begin{pmatrix}
\tilde{h}\left(\bm{k}\right)-\mu & -i\sigma_y\Delta \\
i\sigma_y\Delta & -\tilde{h}^\ast\left(-\bm{k}\right)+\mu
\end{pmatrix}.
\end{align}
Here $\tilde{h}\left(\bm{k}\right)=h\left(\bm{k}\right)+\frac{1}{2}g\mu_{\textrm{B}}\bm{B}\cdot\bm{\sigma}$ is the twelve-band tight binding Hamiltonian matrix that takes into account the Zeeman coupling. The free energy density $F=-\frac{1}{\beta\Omega}\log\mathcal{Z}$ gives Eq. \ref{BdG_free} with $\xi_{\nu,\bm{k}}$ being the eigenvalues of $h_{\textrm{BdG}}\left(\bm{k}\right)$ in Eq. \ref{h_BdG}. The self-consistent mean field equation of the pairing order parameter is given by $\frac{\partial F}{\partial \Delta}=0$. In the calculation, at $B=0\textrm{T}$, we set $T_{\textrm{c}}=7.6\textrm{K}$ and then the effective attraction $U$ can be determined by solving $\frac{\partial F}{\partial \Delta}=0$. By minimizing the free energy density at given $T$ and in-plane magnetic field, the phase diagram of the superconducting pairing state can be obtained.

\section*{Acknowledgment}
The authors are grateful for the helpful discussions with Patrick A. Lee. W.-Y. H. acknowledges the support from the National Natural Science Foundation of China (No. 12304200), the BHYJRC Program from the Ministry of Education of China (No. SPST-RC-10), and the start-up funding from ShanghaiTech University. Q. W. acknowledges the support from the National Key R\&D Program of China (Grant No. 2023YFA1607400).

\onecolumngrid
\clearpage
\begin{center}
{\bf Supplementary Material: Topological Superconductivity in Monolayer T$_d$-MoTe$_2$}
\end{center}

\maketitle
\setcounter{equation}{0}
\setcounter{figure}{0}
\setcounter{table}{0}
\setcounter{page}{1}
\makeatletter

\maketitle
\setcounter{equation}{0}
\setcounter{figure}{0}
\setcounter{table}{0}
\setcounter{page}{1}
\makeatletter
\renewcommand{\figurename}{Supplementary Fig.}
\renewcommand{\tablename}{Supplementary Table}
\renewcommand\thetable{1}

\section*{Supplementary Note 1: The Band Structures and Spin Textures under Different Electric Fields}
In the presence of an out of plane electric field at $E_z$=0.05, 0.1, and 0.15 V/\AA~, we perform the first principle DFT calculations and use the Wannier 90 package to obtain the band structures as depicted in Supplementary Fig. \ref{supfig1}. The out of plane electric field breaks the inversion symmetry and induces the spin splittings in the bands. Different values of $E_z$ slightly changes the spin splittings in the bands, but the spin textures of the Fermi pockets can all be described by the effective Hamiltonian in Eq. 3 in the main text. The monolayer T$_{\textrm{d}}$-MoTe$_2$ with a gate induced $z$-directional electric field belongs to the C$_{1v}$ group, so all the spin textures manifest the unique \emph{Ising plus in-plane SOC} in the monolayer T$_{\textrm{d}}$-MoTe$_2$. Importantly, comparing the spin textures in Fig. \ref{supfig1} (e) with those in Fig. \ref{supfig1} (d) and (f), one can identify a sign change in the $x$ component of $\bm{g}_{\bm{k}}=\left(\lambda_xk_y,\lambda_yk_x,\lambda_zk_y\right)$. The sign change of $g_{x,\bm{k}}$ induced by the $E_z$ field indicates that the $x$ component of $\bm{g}_{\bm{k}}=\left(\lambda_xk_y,\lambda_yk_x,\lambda_zk_y\right)$ can be reduced to zero by changing the value of $E_z$.

\begin{figure}[h]
\centering
\includegraphics[width=0.9\columnwidth]{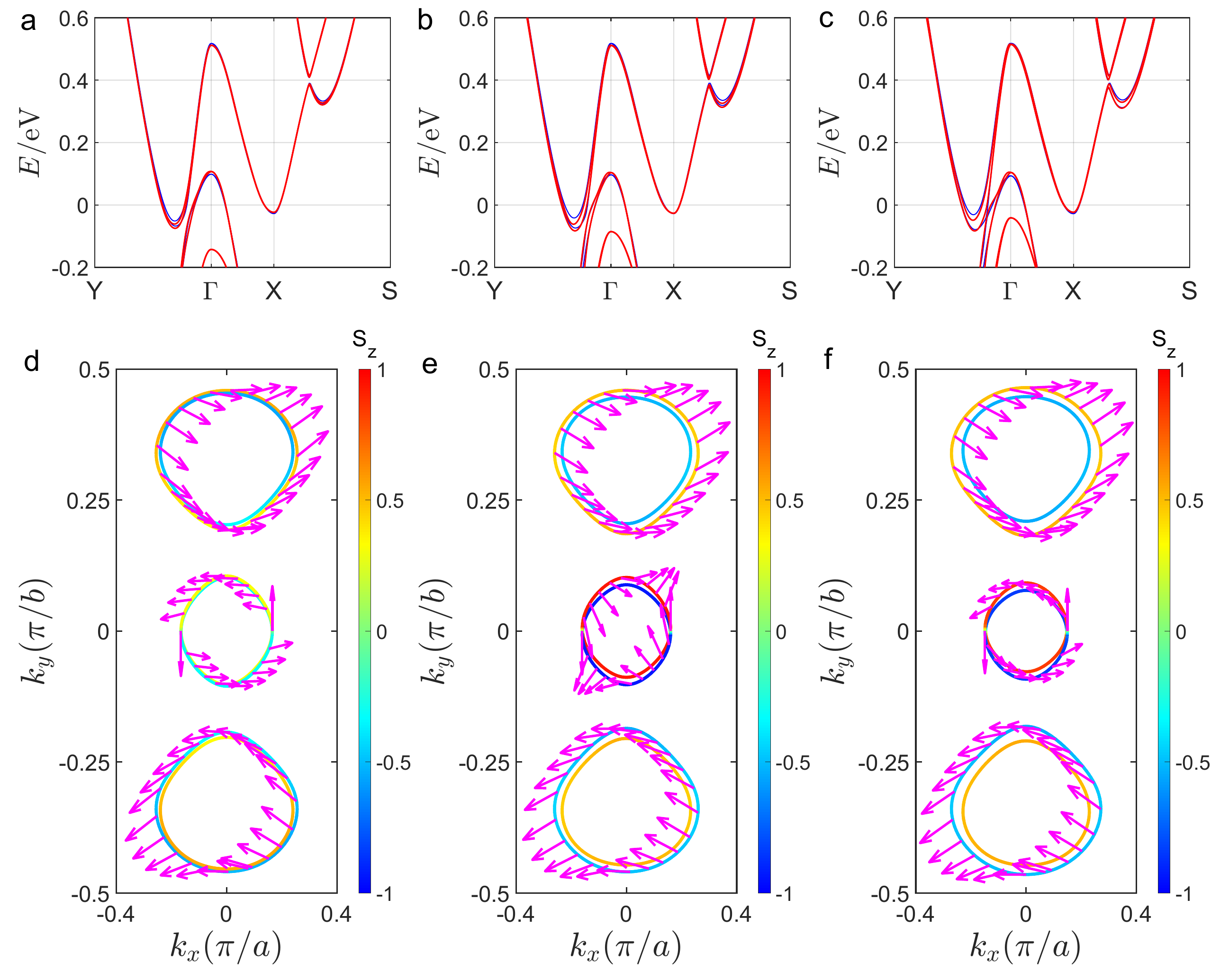}
\caption{The band structure of monolayer T$_{\textrm{d}}$-MoTe$_2$ calculated at $E_z=0.05 \textrm{ V/\AA}$ for (a), $E_z=0.1 \textrm{ V/\AA}$ for (b) and $E_z=0.15 \textrm{ V/\AA}$ for (c). The corresponding spin textures in the Fermi pockets are calculated through the twelve-band effective tight binding model. The results are present in (d), (e) and (f) respectively.}
    \label{supfig1}
\end{figure}

\section*{Supplementary Note 2: The High Tunability of Dual-gate Device}
The dual-gate setup of monolayer T$_{\textrm{d}}$-MoTe$_2$ is schematically plotted in Supplementary Fig. \ref{supfig2} (a). One can regard the dual-gate setup as two quantum capacitors in series. Through tuning the top and bottom gate, one can tune the charge density $n_{\textrm{2D}}$ in the monolayer T$_{\textrm{d}}$-MoTe$_2$ and the effective $z$-directional displacement field individually as~\cite{gatezheng, gatecao}:
\begin{align}
n_{\rm{2D}}=\frac{\epsilon_0 \epsilon_b V_{bg}}{ed_b}+\frac{\epsilon_0 \epsilon_t V_{tg}}{ed_t},\quad\quad D_{z}=\frac{1}{2}\left(\frac{\epsilon_0 \epsilon_b V_{bg}}{d_b}-\frac{\epsilon_0 \epsilon_t V_{tg}}{d_t}\right),
\end{align}
where $\epsilon_{b/t}$ denotes the relative perimitivity in the bottom/top region, and the distances $d_{b/t}$ are denoted in Fig. \ref{supfig2} (a). We know that the chemical potential of a two-dimensional electronic system is determined by the electron density. We also know that a $D_z$ field creates an electric field $E_z=D_z/\epsilon_0$. As a result, the dual-gate setup can tune the chemical potential and $E_z$ field individually by tuning the top and bottom gate voltage. Given $E_z=0.1 \textrm{ V/\AA}$, the in-plane magnetic field induced Zeeman coupling opens a Zeeman gap at $\Gamma$ as shown in Supplementary Fig. \ref{supfig2} (b) and (c). If the chemical potential is tuned into the Zeeman gap at $\Gamma$, the electronic DOS gets reduced and manifests as a dip as shown in Fig. \ref{supfig2} (c). Such in-plane Zeeman coupling induced dip in the electronic DOS is expected to get identified through a magneto-conductance measurement. 

In the experimental accessible range of {$E_z\in\left[-0.1, 0.1\right]$ V/\AA}, we calculate the band structures of the monolayer T$_{\textrm{d}}$-MoTe$_2$ and fit them by the effective hole pocket Hamiltonian around the $\Gamma$ in Eq. 3. The strength of the \emph{Ising plus in-plane SOC} as a function of the applied $E_z$ field is plotted in Supplementary Fig. \ref{supfig3}. It shows that the SOC field $\bm{g}_{\bm{k}}=(\lambda_xk_y,\lambda_yk_x,\lambda_zk_y)$ can be modulated by varying the $E_z$ field. Importantly, due to $\bm{g}_{\bm{k}}\propto\bm{E}\left(\bm{r}\right)\times\left(\hbar\bm{k}+\hat{\bm{p}}\right)$, the $x$ component of the SOC $g_{x,\bm{k}}=\lambda_xk_y$ is greatly affected by the applied $E_z$ field. Within the range of $E_z\in \pm[0.05,0.07]$ V/\AA, $\lambda_x$ is found to reverse its sign, while $\lambda_y$ and $\lambda_z$ keep increasing with $E_z$. Our calculation results indicate that in the dual-gate device, tuning the $E_z$ field can reduce the $x$ component of $\bm{g}_{\bm{k}}$ and make $\bm{g}_{\bm{k}}$ more favorable to the in-plane magnetic field driven TSC.

\begin{figure}[h]
    \centering
    \includegraphics[width=0.9\columnwidth]{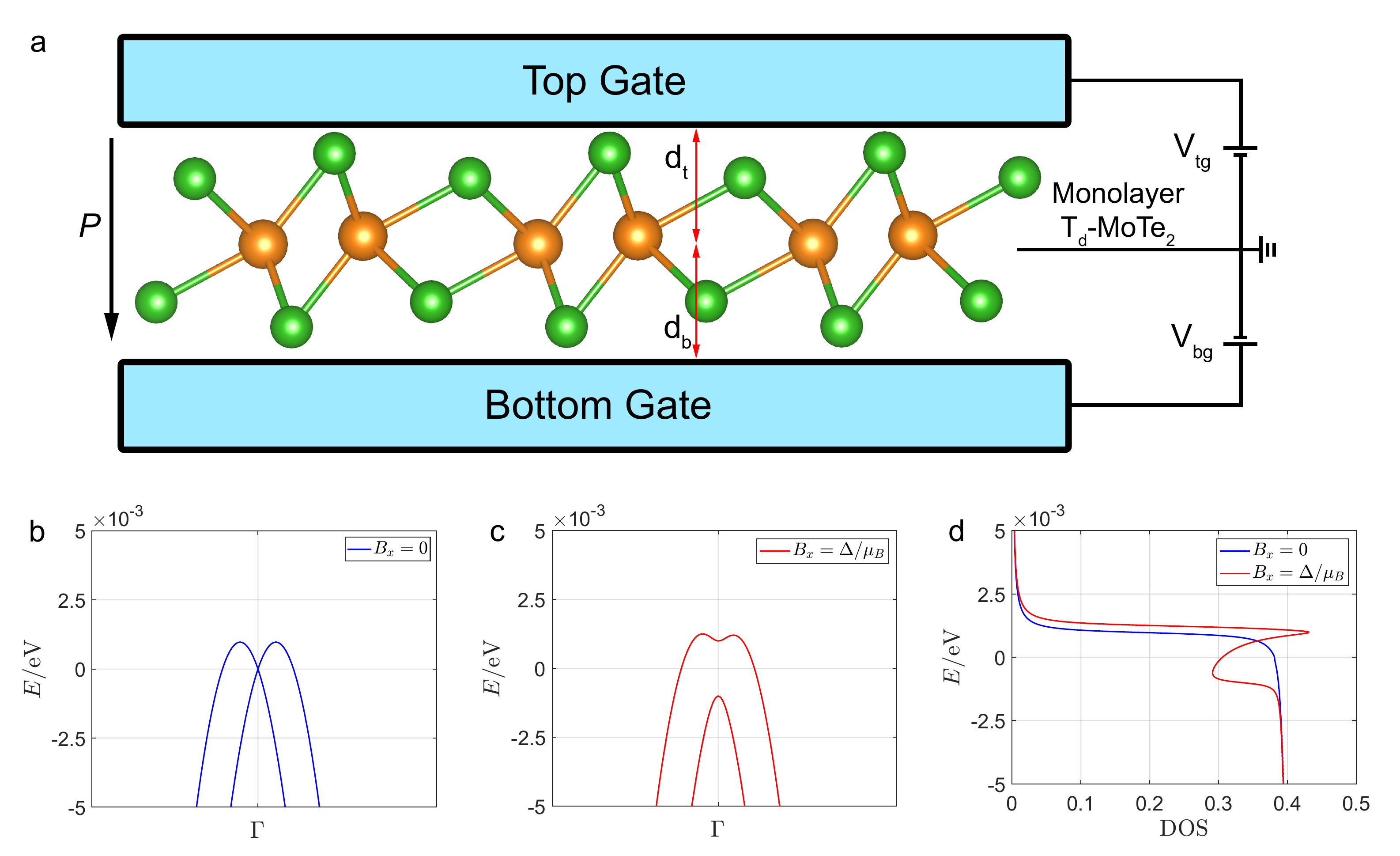}
    \caption[short]{ (a) The schematic showing of a dual-gate device with a monolayer T$_{\textrm{d}}$-MoTe$_2$ crystal involved. (b) and (c) demonstrate the band structure around the $\Gamma$ without/with an in-plane magnetic field. Here we consider the applied electric field in the calculation to be $E_z=0.1$ V/\AA. (d) The electronic DOS for the band structures in (b) and (c). An in-plane magnetic field introduces a Zeeman gap at $\Gamma$ and reduces the DOS.}
    \label{supfig2}
\end{figure}
\begin{figure}[h]
    \centering
    \includegraphics[width=0.5\columnwidth]{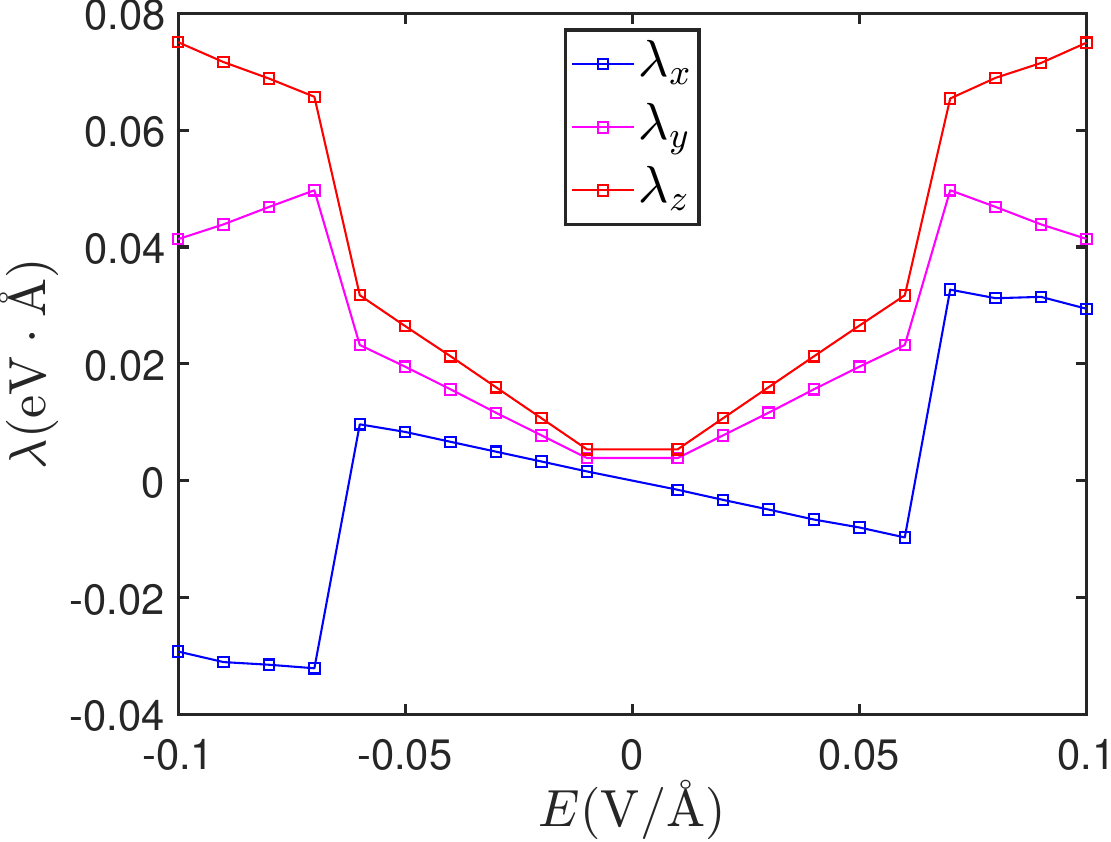}
    \caption[short]{The strength of the \emph{Ising plus in-plane SOC} as a function of the applied $E_z$ field. Within the range of {$E_z\in\pm\left[0.05,0.07\right]$ V/\AA}, the $x$ component of $\bm{g}_{\bm{k}}$ is shown to get reduced.}
    \label{supfig3}
\end{figure}

\section*{Supplementary Note 3: The Calculations for the Density of States and the Energy Spectrum at the Edge}
The electronic DOS in the bulk of the superconducting monolayer T$_{\textrm{d}}$-MoTe$_2$ is given by 
\begin{align}
N\left(E\right)=-\frac{1}{\pi}\textrm{Im}G^{\textrm{R}}\left(E\right).
\end{align}
Here the regarded Green's function $G^{\textrm{R}}\left(E\right)$ can be derived as:
\begin{align}
G^{\textrm{R}}\left(E\right)=\frac{1}{\Omega}\sum_{\bm{k}}\frac{1}{E+i0^+-h_{\textrm{BdG}}\left(\bm{k}\right)},
\end{align}
which is obtained through the analytic continuation of $G\left(i\omega_n,\bm{k}\right)$. In the basis of $\left[c^\dagger_{\bm{k},o,\uparrow}, c^\dagger_{\bm{k},o \downarrow},c_{-\bm{k},o,\uparrow}, c_{-\bm{k},o,\downarrow}\right]^{\textrm{T}}$, the specific form of the BdG Hamiltonian for the monolayer superconducting T$_{\textrm{d}}$-MoTe$_2$ with an in-plane magnetic field reads
\begin{align}
h_{\textrm{BdG}}\left(\bm{k}\right)=&\begin{pmatrix}
h\left(\bm{k}\right)-\mu I_{12}+\frac{1}{2}g\mu_{\textrm{B}}\bm{B}\cdot\bm{\sigma}\otimes I_{6} & -i\sigma_y\Delta\otimes I_{6} \\
i\sigma_y\Delta\otimes I_{6} & -h^\ast\left(-\bm{k}\right)+\mu I_{12}-\frac{1}{2}g\mu_{\textrm{B}}\bm{B}\cdot\bm{\sigma}^\ast\otimes I_6
\end{pmatrix},
\end{align}
with $I_n$ denoting the $n\times n$ identity matrix. Here the in-plane magnetic field takes the form $\bm{B}=B\left(\cos\theta,\sin\theta, 0\right)$. 

By applying the periodic boundary condition solely in the $x$ direction, one can derive the BdG Hamiltonian $h^{\textrm{ribbon}}_{\textrm{BdG}}\left(k_x\right)$ of a monolayer T$_{\textrm{d}}$-MoTe$_2$ ribbon along $x$ through the Fourier transformation. By applying the surface Green's function method~\cite{surfacegf}, one can obtain the retarded Green's function $G^{\textrm{R},\textrm{ribbon}}\left(E,k_x\right)$ at the edge of a monolayer T$_{\textrm{d}}$-MoTe$_2$ ribbon that is semi-infinite in the $y$ direction. The electronic DOS at the edge of the topological superconducting monolayer T$_{\textrm{d}}$-MoTe$_2$ ribbon can then be calculated as
\begin{align}
A\left(E, k_x\right)=-\frac{1}{\pi}\textrm{Im}G^{\textrm{R},\textrm{ribbon}}\left(E,k_x\right).
\end{align}
Performing a summation over $k_x$ gives the electronic DOS at the edge:
\begin{align}
N_{\textrm{edge}}\left(E\right)=\frac{1}{L}\sum_{k_x}A\left(E,k_x\right),
\end{align}
with $L$ denoting the length of the ribbon. The result of $A\left(E, k_x\right)$ and $N_{\textrm{edge}}\left(E\right)$ at $B=1.1\Delta/\mu_{\textrm{B}}$ is plotted in Supplementary Fig. \ref{supfig4}. It is clear that the chiral Majorana edge state contributes to a plateau in the electronic DOS at the energy within gap of the BdG quasiparticle spectrum. 

\begin{figure}[h]
    \centering
    \includegraphics[width=0.5\columnwidth]{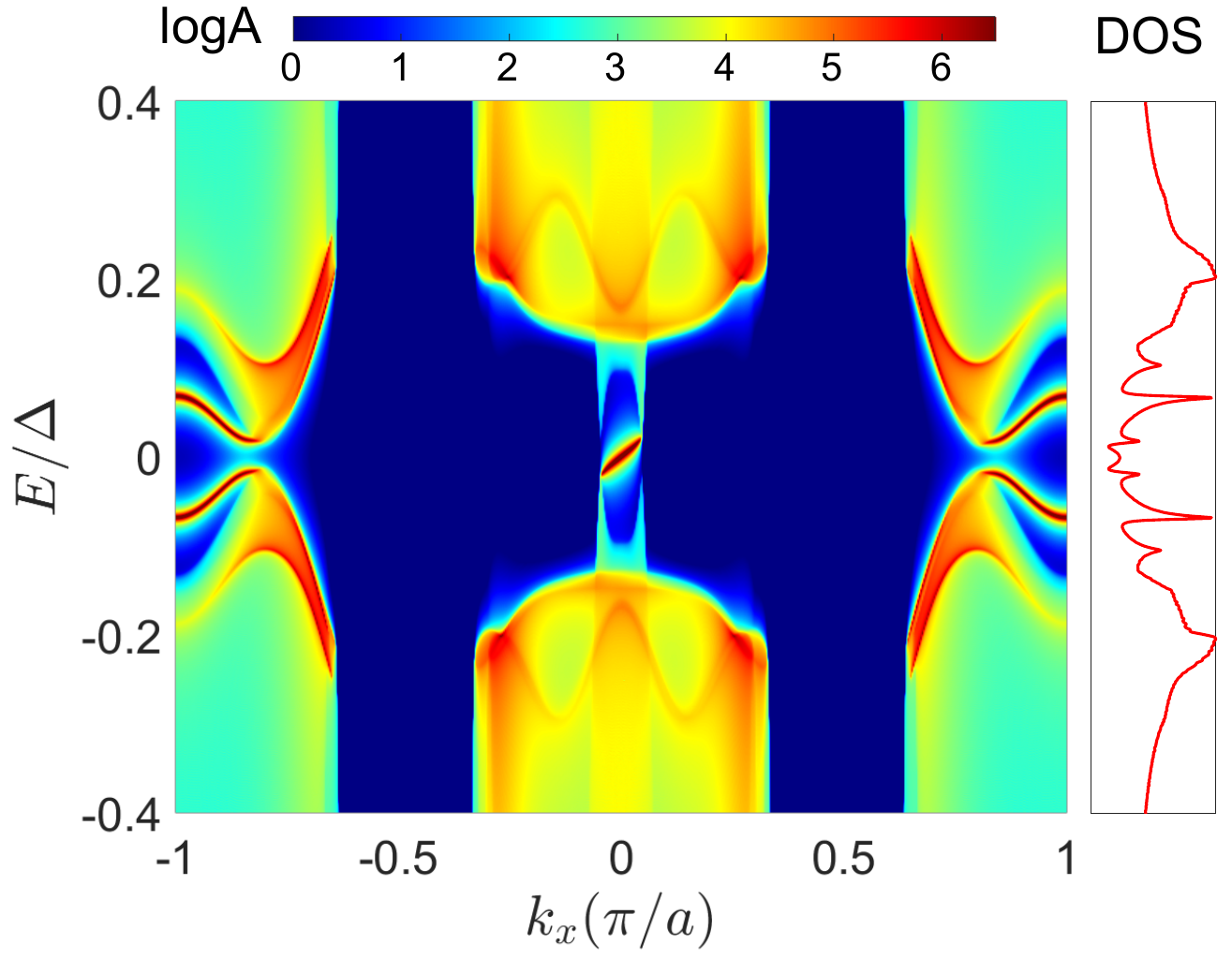}
    \caption[short]{The electronic DOS at the edge of the topological superconducting monolayer T$_{\textrm{d}}$-MoTe$_2$ ribbon. The ribbon respects the periodic boundary condition along $x$ and is semi-infinite along $y$. The left panel is $A\left(E,k_x\right)$ and the right panel corresponds to $N_{\textrm{edge}}\left(E\right)=\frac{1}{L}\sum_{k_x}A\left(E,k_x\right)$. The in-plane magnetic field applied is set to be $B=1.1\Delta/\mu_{\textrm{B}}$.}
    \label{supfig4}
\end{figure}

In the main text, we mainly focus on the in-plane magnetic field applied along the $x$ direction. In Supplementary Fig. \ref{supfig5}, we show how the BdG quasiparticle energy spectrum evolve as the direction of the applied in-plane magnetic field deviates from the $x$ direction. It is clear that the whole spectrum get tilted in the $k_x$ direction. As long as the bulk gap does not close, the topology of the superconducting state does not change. It indicates that our proposal of the in-plane magnetic field driven topological superconducting state in the monolayer T$_{\textrm{d}}$-MoTe$_2$ is not sensitive to the direction of the applied in-plane magnetic field.

\begin{figure}[h]
    \centering
    \includegraphics[width=\columnwidth]{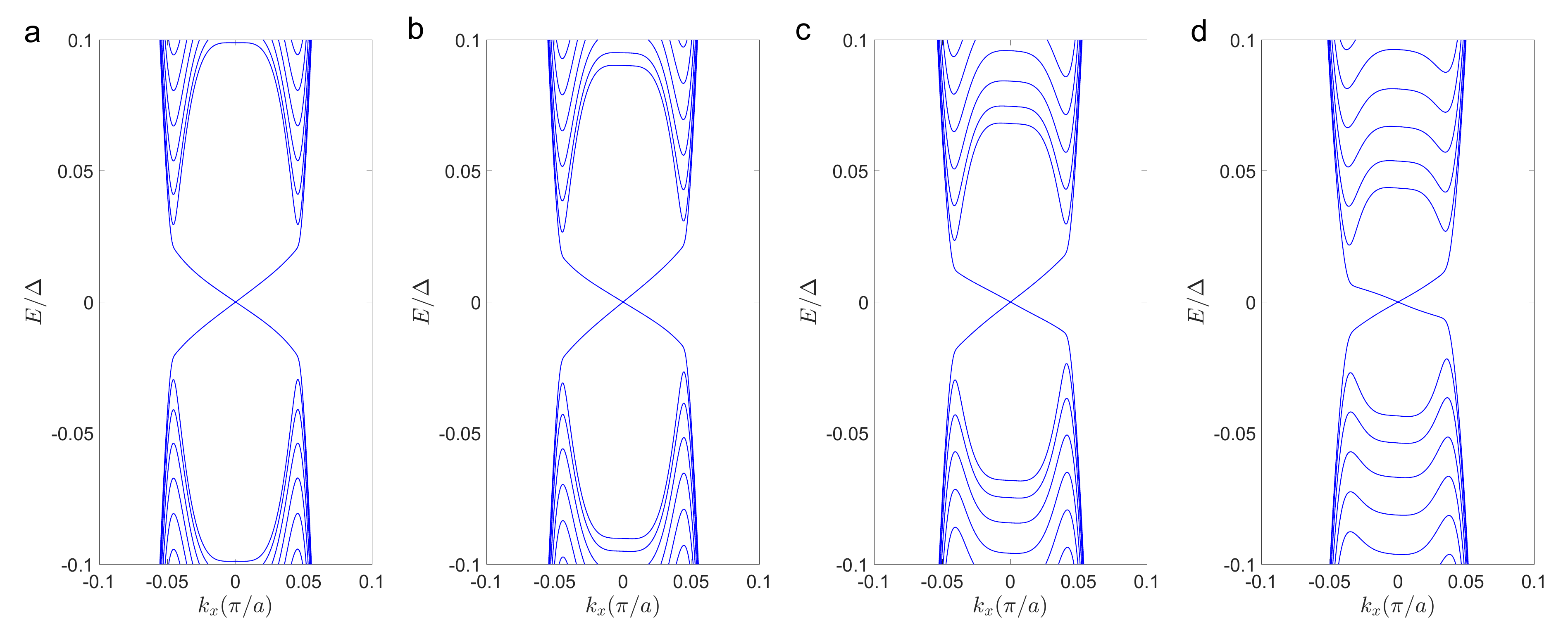}
    \caption[short]{The BdG quasiparticle energy spectrum with a magnetic field $B=1.1\Delta/\mu_{\textrm{B}}$ along different in-plane directions:  (a) $\theta=0^\circ$, (b) $\theta=20^\circ$, (c) $\theta=40^\circ$ and (d) $\theta=60^\circ$. Here $\theta$ is the angle between the direction of the in-plane magnetic field and the positive direction of the $x$-axis.}
    \label{supfig5}
\end{figure}

\end{document}